\documentclass[preprint,prb]{revtex4}
\usepackage{epsfig,graphics,bm}

\newcommand{\PbEuTe}{Pb$_{0.1}$Eu$_{0.9}$Te}
\newcommand{\sigmam}{$\sigma^-$}

\newcommand{\fd}{$4f^7\rightarrow 4f^65d \left(t_{2g}\right)$}
\newcommand{\Euion}{Eu$^{2+}$}
\newcommand{\ket}[1]{|#1\rangle}

\begin{document}
\title{Sharp lines in the absorption edge of EuTe and \PbEuTe\ in high magnetic fields}
\author{L.K.Hanamoto, A.B.Henriques, and N.F.Oliveira, Jr. }
\affiliation{Instituto de F\'{\i}sica, Universidade de S\~ao Paulo\\
Caixa Postal 66318, 05315-970 S\~ao~Paulo, Brazil}
\author{P.Rappl, E.Abramof, and Y.Ueta}
\affiliation{LAS-INPE, Av. dos Astronautas, 1758 - Jd. Granja,
12227-010, S\~ao Jos\'e dos Campos, Brazil}
\begin{abstract}
The optical absorption spectra in the region of the \fd\ transition energies
of epitaxial layers of of EuTe and \PbEuTe, grown by molecular beam epitaxy, 
were studied using circularly polarized light, in the Faraday configuration.
Under \sigmam\ polarization a sharp symmetric absorption line (full width at half-maximum 0.041~eV)
emerges at the low energy side of the band-edge 
absorption, for magnetic fields intensities greater than 6~T. 
The absorption line shows a huge red shift (35~meV/T)
with increasing magnetic fields.
The peak position of the absorption line 
as a function of magnetic field is dominated by the {\em d-f} exchange interaction
of the excited electron and the \Euion\ spins in the lattice.
The {\em d-f} exchange interaction energy was estimated to be $J_{df}S=0.15\pm 0.01$~eV.
In \PbEuTe\ the same absorption line is detected, but it is broader, due to alloy disorder, 
indicating that the excitation is localized within a finite radius.
From a comparison of the absorption spectra in EuTe and \PbEuTe\, 
the characteristic radius of the excitation is estimated to be $\sim 10$\AA.
\end{abstract}
%76.40=cyclotron resonance;72.20=electrical conductivity in semiconductors
%and insulators; 73.20=electron gas 2D; 73.40=electronic transport in interface
%structures; 71.18=Fermi surfaces;
%73.61=low dimensional quantum structures-electrical properties
%72.20.M=Magnetoresistance in semiconductors
%73.61.E=III-V Semiconductors-thin film and layered structures-electrical properties
%73.20.D=Superlattices-electron states
%\author{PACS 73.61.-r, 73.20.-r, 73.20.Dx, 72.20.My}
\maketitle
\section{Introduction}
Europium
chalcogenides are prospective candidates to serve as 'spin-filters' in prototype spintronic devices
\cite{divincenzo} and novel magneto-optical devices (tunable lasers, detectors and modulators) \cite{heissapl},
where the principal source of the interesting physical properties is the
{\em d-f} exchange interaction between band-edge electrons and localized \Euion\ spins.
The optical properties of europium chalcogenides have already been investigated 
some 20-30 years ago, 
and it was established that the band-edge absorption spectrum is associated with
a $4f^7\rightarrow 4f^65d$
electronic transition (surveys are given in Refs. \cite{kasuyarmp,wachter,mauger,schoenes}). 
The experiments were performed on 
powders \cite{wachterSSC70}, polycrystalline thin layers \cite{wachterCRSSS}
and bulk crystals \cite{pidgeon,akimotoJPSJ}, whereby the \fd\ absorption
spectrum is characterized by   
a very broad band - full width at half maximum of almost $\sim$1~eV.
This absorption band was analyzed by some researchers \cite{kasuyarmp,kasuyajmmm} in terms of 
an excitonic transition extending mainly in the nearest neighbor Eu sites, with an exciton  
binding energy dependence on external fields and temperature that is 
governed by the {\em d-f} exchange interaction between the electron and the
\Euion\ spins in the lattice (the magnetic exciton model).
The large breadth of the absorption line, however,
has remained unexplained \cite{kasuyajmmm}, and it  
has lead some researches to propose, instead, that the absorption spectrum 
is associated with a transition to extended conduction band states \cite{wachter,pidgeon,mauger}. 
The photoluminescence of europium chalcogenides was also studied,
and it is described by a broad line (full width at half maximum FWHM=150~meV in EuTe) with a large Stokes shift 
(photon energy $h\nu=1.48$~eV in EuTe) \cite{akimotoJPSJ}.

Recently, molecular beam epitaxy
(MBE) was used to produce EuTe,
which has lead to the observation of a much narrower photoluminescence line 
(full width at half maximum FWHM=10~meV), 
and at a much photon higher energy ($h\nu=1.92$~eV) \cite{heissapl,heiss} than in previous
investigations. 
The photoluminescence line displays a giant red shift (34~meV/T) when a magnetic field
is applied. 
This sensitivity was interpreted in the framework of the 
magnetic exciton model \cite{kasuyarmp,kasuyajmmm}. 
The theory relies on such 
basic parameters as the {\em d-f} exchange interaction energy and effective Bohr radius, whose accepted values
vary within a large interval. 
For example, the {\em d-f} exchange interaction energy, $J_{d-f}S$ in EuTe, 
has been estimated to be (a) 0.35~eV, based on measurements on the temperature dependence
of the optical absorption edge \cite{wachter,umehara95}, (b) 0.11~eV from the red shift of the Faraday rotation
spectrum, and also of the optical band-edge, when a high magnetic field is applied \cite{schoenes79},
(c) 0.17~eV from theoretical calculations \cite{cho,umehara02}.

In this work we studied the \fd\ optical 
absorption of epitaxial layers of EuTe grown by MBE, using circularly polarized light and
high magnetic fields. At zero magnetic field, the optical absorption spectrum is described
by a sharp threshold around 2.25~eV. 
When the intensity of the applied magnetic field is increased above 6~T,
a very sharp symmetric
absorption line 
(full width at half maximum of 41~meV) appears at the low energy
side of the absorption onset in one of the photon polarizations ($\sigma^-$).
With increasing magnetic field
this absorption line displays a huge red shift,
which saturates when the \Euion\ spins in the crystal arrange ferromagnetically.
The peak position dependence on the magnetic field fits very well
into the model of a localized excitation whose energy is 
tunable through the {\em d-f}
exchange interaction. From the absorption line shift with magnetic field we
obtained an estimate of the {\em d-f} exchange interaction energy, $J_{d-f}S=0.15$~eV, with an uncertainty
less than 10\%.

The optical absorption dependence on magnetical field was also studied for a Pb$_{0.1}$Eu$_{0.9}$Te sample.
The optcial absorption of Pb$_{1-x}$Eu$_x$Te for $x>0.6$ is EuTe-like, i.e. it is associated with an
$4f^7\rightarrow 4f^65d$ electronic transition, whose energy is pinned at 2.3~eV for $x>0.8$\cite{bauer}.
In the Pb$_{0.1}$Eu$_{0.9}$Te sample, the same absorption line seen in EuTe was detected in high
magnetic fields (above 6~T), however, the maximum red shift of the line was smaller, due to a smaller
concentration on \Euion in the lattice. In addition, the width of the absorption line (43~meV) was larger
than in EuTe (41~meV). Assuming that the increase in the breadth of the line is due to alloy disorder,
the radius of localization of the excitation is estimated to be $\sim$10\AA.

Recent investigations of the photoluminescence spectrum of EuTe grown by MBE have 
discovered almost the same red shift of the optical line (34~meV/T) when a magnetic field is
applied, as in this work. It should be pointed out, however, that the physical mechanisms involved in
the photoluminescence could be much more complex than in the optical absorption.
According to the Franck-Condon principle \cite{condon}, whereas absorption occurs
when the crystal lattice is undistorted, emission occurs when the atoms surrounding
the excitation move to their new equilibrium positions, and in EuTe a large distortion
occurs due to the a large charge transfer from to excited \Euion\ to the surrounding atoms \cite{kasuyarmp};
in addition to the mechanical distortion, an alignment of the \Euion\ spins inside the 
excitation radius (forming a magnetic exciton), 
would also contribute to reduce the energy of the excitation.
The amount by which energy is reduced could be dependent on the intensity of the
magnetic field applied, due to
a changing size of the excitation, making 
the magnetic field induced red shift seen in the PL much more
complex 
than the red shift of the absorption spectrum reported in this work, in which the positions of the
atoms in the lattice, and their spin orientations, are fixed at the unperturbed crystal state.

\section{Experimental}
EuTe and Pb$_{0.1}$Eu$_{0.9}$Te epitaxial layers ($\sim 1.1\mu$m thick) were grown by
molecular beam epitaxy (Riber 32P MBE) onto freshly cleaved (111)
BaF$_2$ substrates. In the case of EuTe, the constituents (Eu and Te) were
sublimated from elemental sources, while a PbTe source was used for the
\PbEuTe\ growth. Pb$_{1-x}$Eu$_x$Te crystallizes in the rock salt structure, and
its lattice mismatch to the BaF$_2$ substrate (fluoride structure) varies from
4 to 6\%, with increasing Eu content in the ternary compound. In addition,
the thermal expansion coefficient of BaF$_2$ is well matched to those of PbTe
and EuTe, thus preventing the build up of high thermal strains when cooling
the samples to cryogenic temperatures. The Eu content values 
in the Pb$_{0.1}$Eu$_{0.9}$Te sample was determined by magnetization SQUID measurements, 
with an estimated uncertainty of less than one percent.
The crystalline quality was investigated by measuring the x-ray rocking
curve of the (222) Bragg reflection of the samples. The peak position
corresponded to the bulk lattice parameter of 6.600\AA, and the full width
at half maximum of the EuTe diffraction peak was about 400 arcsec, which is
indicative of a good structural quality (for \PbEuTe\ the FWHM of the peak 
was larger than in EuTe
by about 100~arcsec, due to alloy disorder). PL measurements at T=2K 
were made, using the 488nm line of an Ar laser, and the sharp luminescence (FWHM$\sim$10~meV)
at 1.922~eV was detected,
demonstrating the high optical quality of the samples \cite{heissapl,heiss}.
The optical absorption
spectra of EuTe samples and \PbEuTe\  were measured at 1.8K, using left
and right circularly polarized light, in magnetic fields of intensity up
to 17 Teslas. The sample (square shaped, of side length 2.5~mm), was
immersed in superfluid helium, and light was conveyed to the sample, and
collected from the sample, in the Faraday configuration, using optical
fibers coupled to {\em in situ} miniature focusing optics and circular
polarizers.

\section{Results}
\begin{figure}[h]
\centerline{\hspace{-1cm}\epsfxsize=8cm\epsffile{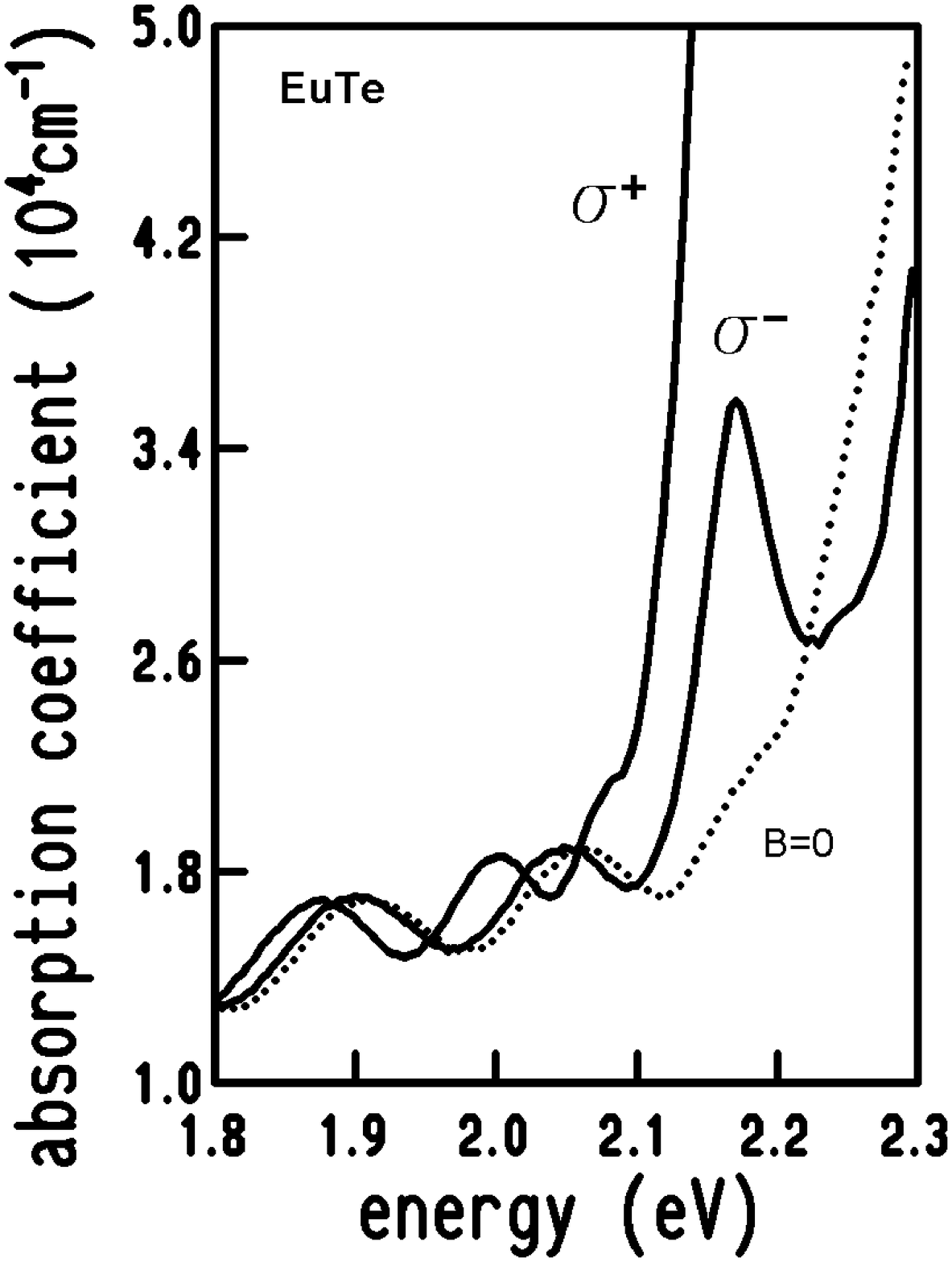}}
\caption{}
\label{fig:fig1}
\end{figure}

The absorption spectrum taken at zero field shows the characteristic absorption onset
at $\sim$2.3 eV (see Fig.\ref{fig:fig1}), which is associated with
the \fd\ transition \cite{wachter,mauger}. In the transparency region, the
absorption coefficient shows weak oscillations, due to the effect of Fabri-Perot
interference in the thin epitaxial film.

When a magnetic field is applied, 
the absorption edges taken in both polarizations show remarkably different behaviours. 
In the $\sigma^+$ polarization there is simply a red shift of the absorption edge,
which saturates when the applied field, $B_{\rm a}$, reaches 8.3~T.
In the $\sigma^-$ polarization, however, the absorption spectrum does not red shift as
a whole, but develops a new sharp absorption peak that emerges from the main absorption
edge. With increasing magnetic field the new absorption peak gains intensity, and it red shifts,
until saturation is reached at $B_{\rm a}$=8.3~T.
The same absorption spectra were measured for a number of EuTe samples of various thickness in the
range of 1-3$\mu$m, some of which contained a
BaF$_2$ protective layer, some of which did not. The position and width of the peak was the same in all samples,
and the optical density of the sharp absorption peak was proportional to the thickness of the EuTe epitaxial
layer, demonstrating that the absorption line is originated in the bulk of EuTe.

For our sample geometry (a thin layer of EuTe subject to a perpendicular
magnetic field), the internal field will be given by $B_{\rm int}=B_{\rm a}-\mu_0 M$, 
where $M$ is the magnetization of the sample, and $\mu_0M$ varies linearly with
$B_{\rm a}$ and saturates to $\sim 1.1$~T at $B_{\rm int}=7.2$~T \cite{oliveira}, 
meaning that $B_{\rm a}$=8.3~T corresponds to an internal field equal to the well known critical field 
value of 7.2~T for EuTe.
We have employed fields as large as 17~T, but no further changes were seen in the absorption spectra.
The absorption spectra at $B_{\rm a}$=9.6~T in both polarizations are shown in Figure~1.

To show the evolution of the absorption spectra taken in the $\sigma^+$ and $\sigma^-$
polarizations when the applied magnetic field is increased from zero, a contour plot of the absorption
spectra is shown in Figure 2. This figure shows that in the $\sigma^+$ polarization the absorption
edge red shifts almost linearly with the field intensity, until saturation is reached. In contrast, the evolution of the absorption
spectrum in the $\sigma^-$ polarization does not show a linear red shift of the absorption edge,
but rather the changes in the spectrum are described by the emergence of a sharp absorption line at
the low energy side of the spectrum. When $B$ is less than the saturation value,
the strength of the sharp absorption line increases with $B$, and it red shifts, leading to a completely resolved
line at the highest fields. 
The red shift of the sharp peak is described by a huge value of
35~meV/T, which is about the same value seen in the shift of excitonic PL \cite{heissapl,heiss}.
Notice also that the main absorption edge remains nearly in the same place
as at $B=0$.

The peak position and full width at half maximum of the
sharp absorption peak seen in the \sigmam\ polarization at fields above 8.3~T 
were obtained from a fit of the absorption spectrum with a linear combination
of a Gaussian curve and a straight line (see Fig.\ref{fig:fig3}). 
For EuTe, the maximum occurs 
at 2.169~eV (the peak   
is shifted by $\sim 0.22$ eV from
the main \fd\ absorption), 
and the full width at half maximum (FWHM) equals to
41~meV. 
The \PbEuTe\ sample showed the same qualitative behavior as seen in EuTe, but the 
low energy side peak was broader than for EuTe, and centered at a 
higher energy, as shown in the right panel of Fig.\ref{fig:fig1}.
The absorption maximum
is seen at 2.181~eV, i.e. an energy 12~meV greater than for EuTe, 
and the FWHM=43~meV, greater by 2~meV compared to
the same value for EuTe (see right panel in Fig.\ref{fig:fig3}).

\section{Discussion}
The magnetic field dependence of the newly discovered absorption line will be discussed in the
framework of the $d-f$ exchange interaction. The 
$d-f$ exchange interaction energy operator is given by \cite{kasuyarmp}
$$
H_{df}=-J_{df} \sum_{n\mu\nu} a^\dagger_{n\mu}a_{n\nu}\bm{S}_n\cdot\bm{\sigma}_{\mu\nu}
$$
where $J_{df}$ is the $d-f$ exchange interaction constant, 
$a^\dagger_{n\mu}$ is the creation operator of a Wannier function
centered at the $n-$th \Euion\ with spin $\mu$, $\bm{\sigma}$ is a Pauli spin operator,
and $\bm{S}_n$ the spin of the $n-$th \Euion.
We treat $H_{df}$ as a perturbation over a photoexcited electron in a $d-$band with definite spin,
whose wave function is given by $\Psi=\sum_n c_n a_{n\chi}^\dagger\ket{0}$.
The spin of the photoexcited electron is subject to the constraint that it must be
parallel to the \Euion\ spin at the site where the excitation occurred \cite{kasuyarmp,umehara95}.
We assume that the wave function describing the excited state encloses a large 
number of \Euion\ sites, hence
at zero magnetic field (T=0K), when the
arrangement of the Eu ions is antiferromagnetic, (the Eu spins lie in the (111) plane \cite{oliveira}), 
the $d-f$ interaction energy of the photoexcited electron averages out to zero. 
As the magnetic field applied normal to the (111) increases, 
the \Euion\ spins cant in the magnetic field direction, the {\em d-f} exchange interaction energy increases
and at arbitrary field it will be given by 
\begin{equation}
\Delta E_{df}=-J_{df}S\cos^2\theta
\label{eq:deltae}
\end{equation}
where $2\theta$ is the angle between the sublattice spin directions, determined by \cite{pidgeon}
$$
%\begin{equation}
\cos\theta=\left\{
\begin{array}{cc} 
\frac{B_{\rm int}}{B_{\rm Sat}} &\mbox{if $B_{\rm int}<B_{\rm Sat}$}\\
1 & \mbox{if $B_{\rm int}>B_{\rm Sat}$} \end{array} 
\right.
\label{eq:costh3}
%\end{equation}
$$
where
$B_{\rm Sat}=7.22~T$ is the saturation field for EuTe \cite{oliveira}. 

Fig.\ref{fig:fig4} shows the measured position of the side peak as a function of applied magnetic field
both for EuTe and \PbEuTe. The same peak position dependence on magnetic field
as for the EuTe sample shown in figure~\ref{fig:fig4} was 
obtained for a second EuTe sample, whose epitaxial layer thickness was 1.75$\mu$m.
The solid lines in Fig.\ref{fig:fig4} were obtained by a simultaneous fit of all of the experimental
values shown to the equation 
\begin{equation}
E=E_0+x\Delta E_{df}
\label{eq:deltaEx}
\end{equation}
where $E_0$ is the position of the excitonic line at $B=0$, 
$x$ is the molar fraction of Eu in the Pb$_{1-x}$Eu$_x$Te
sample, and $\Delta E_{df}$ is given by Eq.(\ref{eq:deltae}). 
Equation (\ref{eq:deltaEx}) should be valid in the range $x>0.8$, where the
Pb$_{1-x}$Eu$_x$Te gap is EuTe-like and nearly constant \cite{bauer}.
In the fitting procedure, the spin of the \Euion\ was fixed at $S=\frac{7}{2}$,
and the $d-f$ exchange interaction
constant, $J_{df}$, and  $E_0$ were the fitting parameters, leading to the best fit values of $E_0$=2.321~eV
and $J_{df}S=0.15\pm 0.01$~eV. 

For \PbEuTe, the energy position of the side peak is 
greater than in EuTe, which is due to a lower density of spins in the lattice. Then,
assuming that in going from EuTe to \PbEuTe,
the increase in the full width at half maximum (FWHM) of the side peak
is due to inhomogeneous broadening, caused
by random fluctuations in the local Eu concentration, the number of \Euion\ spins inside the exciton radius 
can be estimated through
$$
N=\frac{8x(1-x)\left(\Delta E_{\rm max}/\Delta x\right)^2\ln 2}
{
\left(
{\rm FWHM}_{{\rm Pb}_{1-x}{\rm Eu}_x{\rm Te}}
\right)^2
-
\left({\rm FWHM}_{\rm EuTe}\right)^2
}
$$ 
where $\Delta E_{\rm max}$ and $\Delta x$ are the
shift in the position of the maximum of the excitonic line and the change in Eu molar fraction on going from
EuTe to Pb$_{1-x}$Eu$_x$Te, respectively. 
Using the results obtained for EuTe and \PbEuTe,
we find $N=42$, which justifies the assumption made above that the exciton is large, and
means that the exciton sphere encloses about 10 conventional unit cells of the fcc structure.
This estimate puts the exciton diameter at approximately 18\AA. 

In conclusion, we have studied the optical band-edge absorption in 
EuTe and \PbEuTe\ grown by MBE at low temperatures
and high magnetic fields. Under $\sigma^-$ polarization, we detect a well-resolved 
sharp peak (full width 0.041~eV). The sharp absorption line displays a huge red shift of about 35~meV/T 
when the magnetic
field increases, and its separation from the main absorption band reaches $\sim$ 0.2 eV at the saturation field. 
In \PbEuTe, the same absorption line was also detected, but it is broader, which is an indication of a localized
excitation. The displacement of the absorption line with magnetic field fits very well into a
model of a localized excitation 
whose energy is tuned by the {\em d-f} exchange interaction. 
From a fit of the experimental points to the theory we obtain 
a direct estimate of the {\em d-f} exchange interaction energy, $J_{df}S$=0.15$\pm$0.01~eV. 

\section{Acknowledgements} 
We are greatful to Prof. F. Ikawa for help with the photoluminescence measurements, 
done at UNICAMP.
L.K.H. acknowledges support from FAPESP (grant No. 02/00720-9). 
This work was supported by grants CNPq-306335/88-3 and FAPESP-99/10359-7.

\pagebreak
\begin{minipage}[b]{5.6in}
\section*{FIGURE CAPTIONS}
{\bf Fig.~1} Absorption spectra of EuTe for B=0T (dotted line) and B=9.6T (full curves), in the latter case 
the polarization corresponding to each curve ($\sigma^+$ or $\sigma^-$) is indicated.\\ \\

{\bf Fig.~2} Contour plot for the absorption spectrum dependence on the magnetic field intensity (a) $\sigma^-$; 
(b) $\sigma^+$.\\ \\

{\bf Fig.~3} Dots represent the \sigmam\ absorption spectrum of EuTe (left) and \PbEuTe\ (right)
measured at B=9.6T. Full lines are fitting curves to the measured spectra.\\ \\

{\bf Fig.~4} Magnetic field dependence of the energy position of the maximum in the \fd\ excitonic absorption
measured in $\sigma^-$ polarization.\\ \\
\end{minipage}

\pagebreak
\setlength{\unitlength}{1mm}
\vfill
{\tt Manuscript by Henriques {\em et al}, Figure 1}

\pagebreak
\setlength{\unitlength}{1mm}
\begin{figure}[h]
\centerline{\hspace{-1cm}\epsfxsize=8cm\epsffile{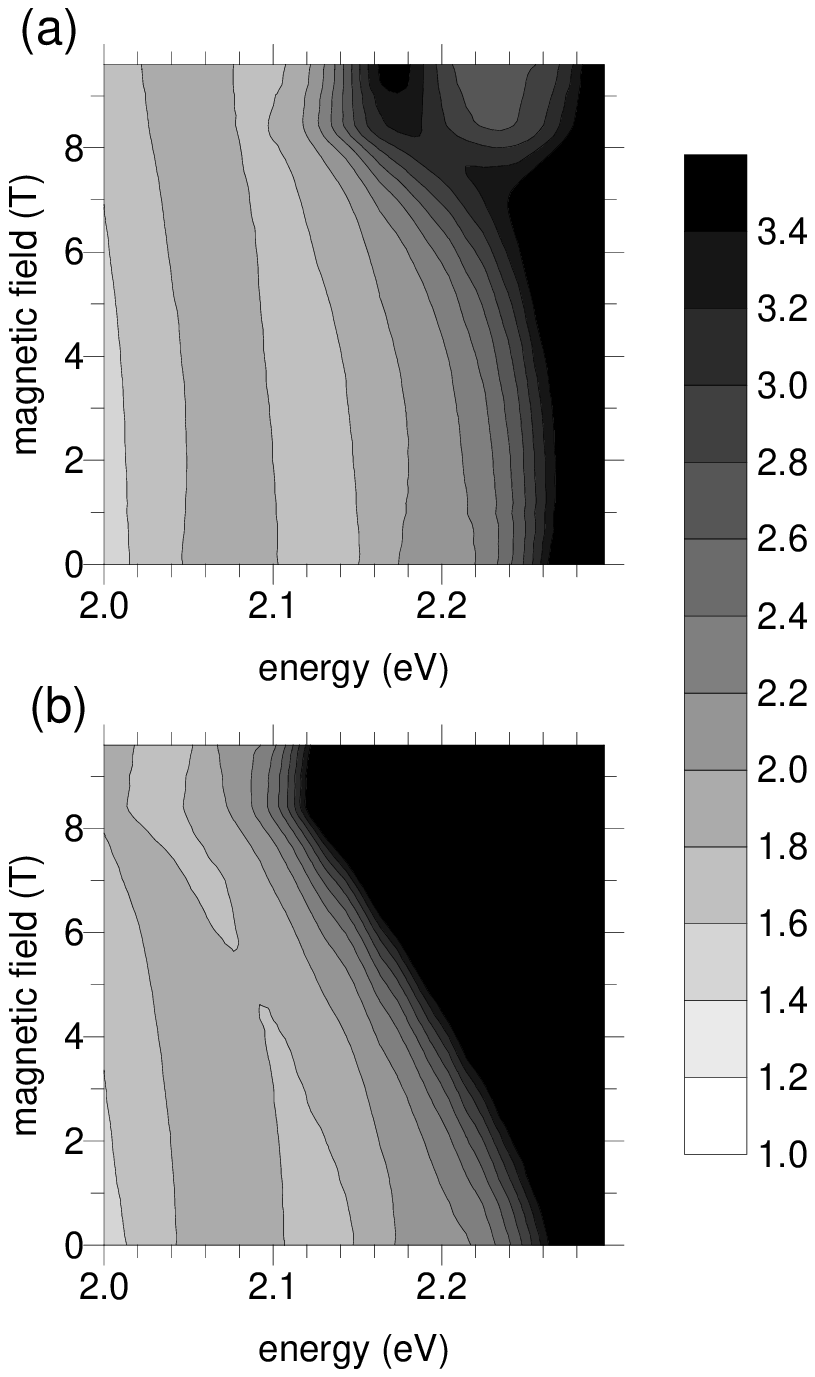}}
\caption{}
\label{fig:fig2}
\end{figure}
\vfill
{\tt Manuscript by Henriques {\em et al}, Figure 2}

\pagebreak
\setlength{\unitlength}{1mm}
\begin{figure}[h]
\centerline{\hspace{-1cm}\epsfxsize=8cm\epsffile{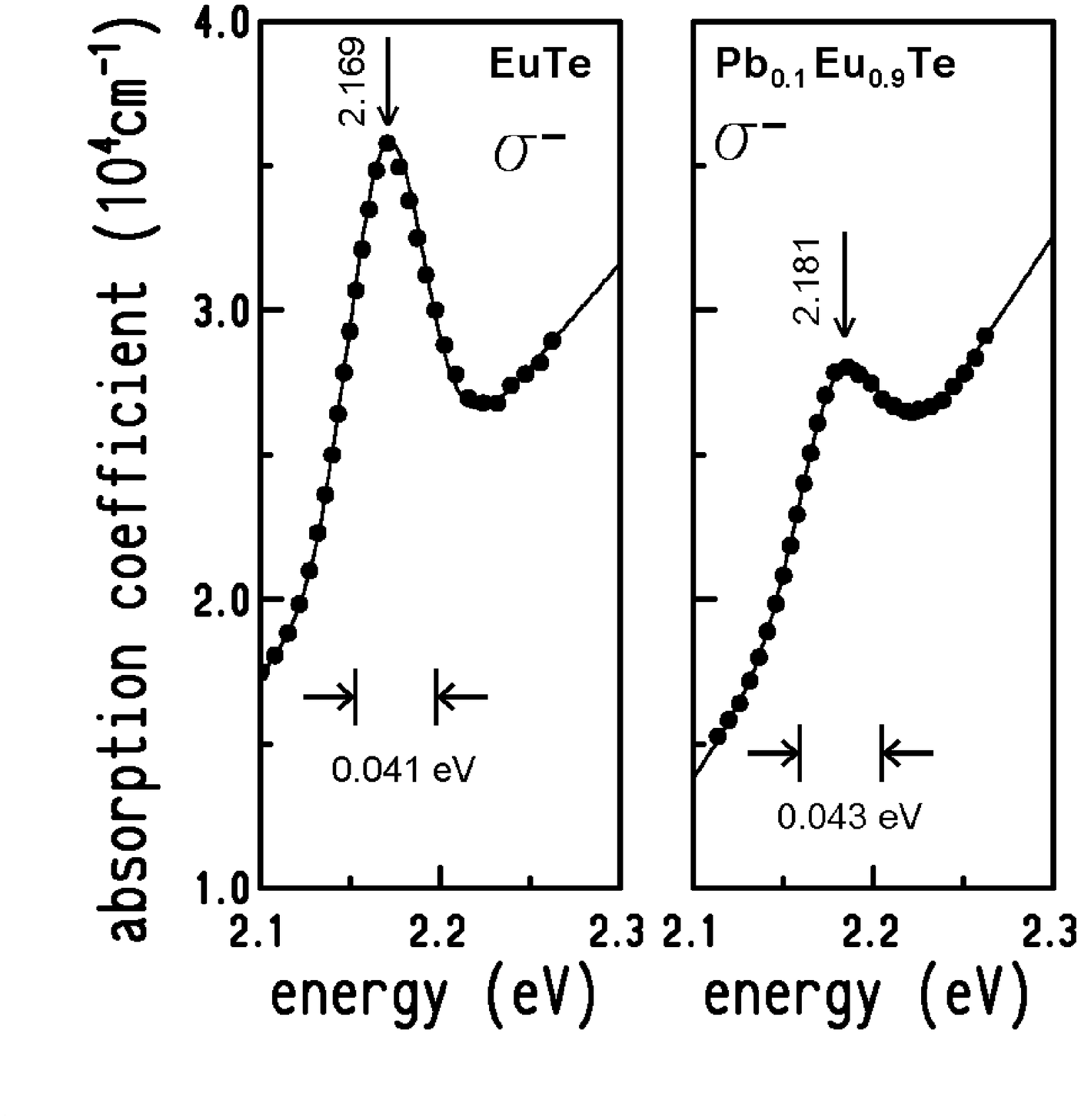}}
\caption{}
\label{fig:fig3}
\end{figure}
\vfill
{\tt Manuscript by Henriques {\em et al}, Figure 3}

\pagebreak
\setlength{\unitlength}{1mm}
\begin{figure}[h]
\centerline{\hspace{-1cm}\epsfxsize=8cm\epsffile{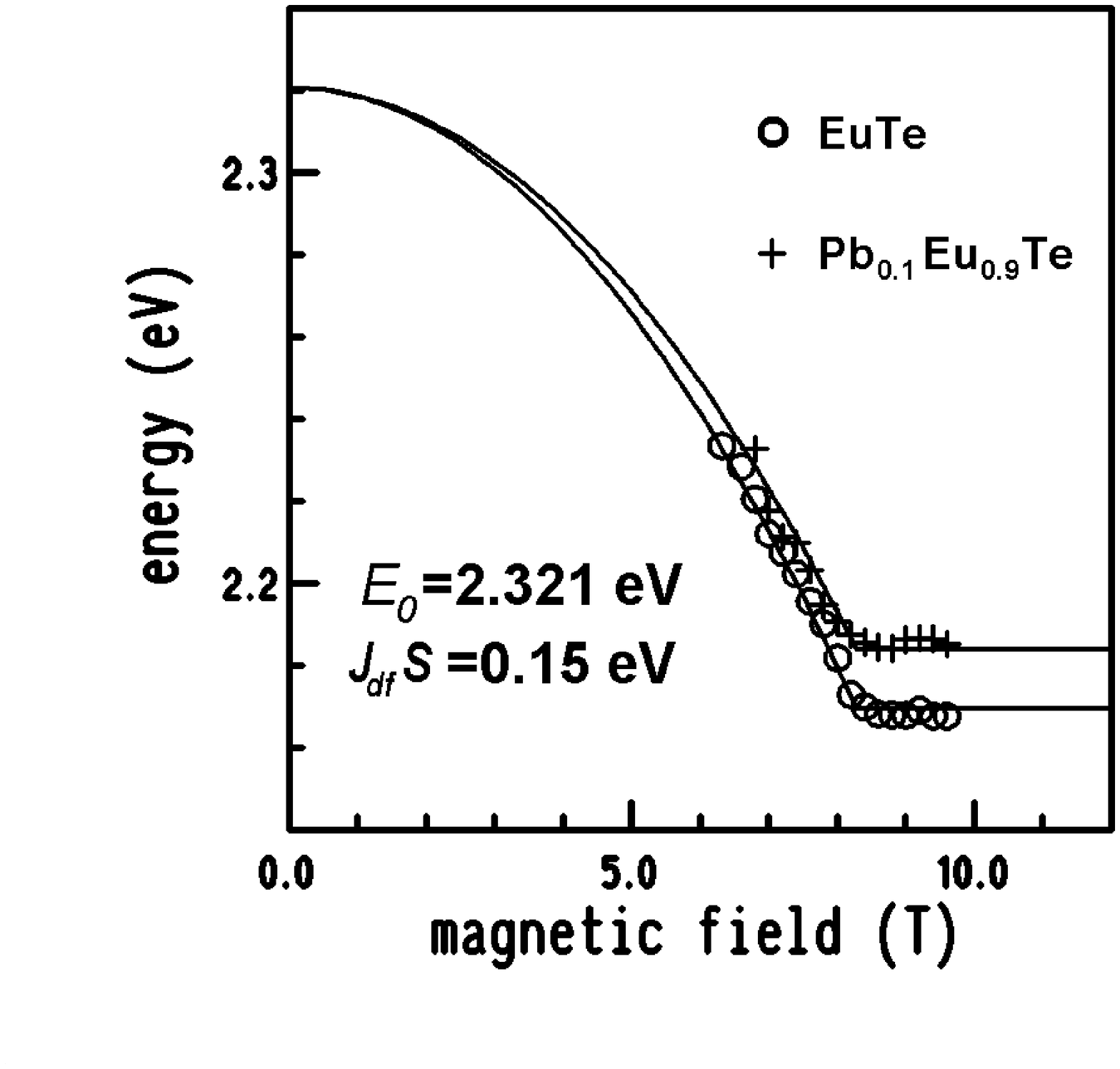}}
\caption{}
\label{fig:fig4}
\end{figure}
\vfill
{\tt Manuscript by Henriques {\em et al}, Figure 4}

\end{document}